\documentclass[]{aa}
\usepackage{graphicx}
\begin{document}

\title{
Temperature and polarization angular power spectra of Galactic dust
radiation at 353~GHz as measured by Archeops}

\subtitle{}

\author{ 
N. Ponthieu \inst{1,\,2} \and  
J. F. Mac\'{\i}as--P\'erez \inst{2} \and  
M. Tristram \inst{2} \and  
P. Ade \inst{3} \and  
A. Amblard \inst{4} \and  
R. Ansari \inst{5} \and  
J. Aumont \inst{2} \and
{\'E}. Aubourg \inst{6, \, 7} \and  
A. Beno\^{\i}t\inst{8} \and  
J.--Ph. Bernard \inst{9} \and  
A. Blanchard\inst{10} \and  
J. J. Bock \inst{11,\,12} \and  
F. R. Bouchet \inst{13} \and  
A. Bourrachot \inst{5} \and  
P. Camus \inst{8} \and  
J.-F. Cardoso \inst{14, \, 7} \and
F. Couchot \inst{5} \and  
P. de Bernardis \inst{15} \and  
J. Delabrouille \inst{14, \, 7} \and  
F.--X. D\'esert \inst{16} \and  
M. Douspis \inst{10} \and  
L. Dumoulin \inst{17} \and  
Ph. Filliatre \inst{18} \and  
P. Fosalba \inst{19} \and
M. Giard \inst{9} \and
Y. Giraud--H\'eraud \inst{14, \, 7} \and  
R. Gispert \inst{1\dag}\thanks{Richard Gispert passed away few weeks
after his return from the early mission to Trapani} \and  
J. Grain \inst{2} \and
L. Guglielmi \inst{14, \, 7} \and  
J.--Ch. Hamilton \inst{20} \and  
S. Hanany \inst{21} \and  
S. Henrot--Versill\'e \inst{5} \and  
J. Kaplan \inst{14, \, 7} \and  
G. Lagache \inst{1} \and  
A. E. Lange \inst{12} \and  
K. Madet \inst{8} \and  
B. Maffei \inst{3} \and  
S. Masi \inst{15} \and  
F. Mayet \inst{2} \and
F. Nati \inst{15} \and  
G. Patanchon \inst{22} \and
O. Perdereau \inst{5} \and  
S. Plaszczynski \inst{5} \and 
M. Piat \inst{14, \, 7} \and
S. Prunet   \inst{13} \and
J.--L. Puget \inst{1} \and  
C. Renault \inst{2} \and  
C. Rosset \inst{14, \, 7} \and  
D. Santos \inst{2} \and  
D. Vibert \inst{13} \and  
D. Yvon \inst{6, \, 7}
}

   \offprints{reprints@archeops.org}
   \mail{macias@in2p3.fr}
\institute{
Institut d'Astrophysique Spatiale, B\^at. 121, Universit\'e Paris
XI, 91405 Orsay Cedex, France
\and
Laboratoire de Physique Subatomique et de Cosmologie, 53 Avenue des 
Martyrs, 38026
Grenoble Cedex, France
\and
Cardiff University, Physics Department, PO Box 913, 5, The Parade,  
Cardiff, CF24 3YB, UK 
\and
University of California, Berkeley, Dept. of Astronomy, 601
Campbell Hall, Berkeley, CA 94720-3411, U.S.A.
\and
Laboratoire de l'Acc\'el\'erateur Lin\'eaire, BP 34, Campus
Orsay, 91898 Orsay Cedex, France
\and
CEA-CE Saclay, DAPNIA, Service de Physique des Particules,
Bat 141, F-91191 Gif sur Yvette Cedex, France
\and
F\'ed\'eration de Recherche APC, Universit\'e Paris 7, Paris, France
\and
Centre de Recherche sur les Tr\`es Basses Temp\'eratures,
BP166, 38042 Grenoble Cedex 9, France
\and
Centre d'\'Etude Spatiale des Rayonnements,
BP 4346, 31028 Toulouse Cedex 4, France
\and
Laboratoire d'Astrophysique de l'Obs. Midi-Pyr\'en\'ees,
14 Avenue E. Belin, 31400 Toulouse, France
\and
California Institute of Technology, 105-24 Caltech, 1201 East
California Blvd, Pasadena CA 91125, USA
\and
Jet Propulsion Laboratory, 4800 Oak Grove Drive, Pasadena,
California 91109, USA
\and
Institut d'Astrophysique de Paris, 98bis, Boulevard Arago, 75014 Paris,
France
\and
Physique Corpusculaire et Cosmologie, Coll\`ege de
France, 11 pl. M. Berthelot, F-75231 Paris Cedex 5, France
\and
Gruppo di Cosmologia Sperimentale, Dipart. di Fisica, Univ. ``La
Sapienza'', P. A. Moro, 2, 00185 Roma, Italy
\and
Laboratoire d'Astrophysique, Obs. de Grenoble, BP 53,
38041 Grenoble Cedex 9, France
\and
CSNSM--IN2P3, Bât 108, 91405 Orsay Campus, France
\and
CEA-CE Saclay, DAPNIA, Service d'Astrophysique, Bat 709,
F-91191 Gif sur Yvette Cedex, France
\and
Institute for Astronomy, University of Hawaii, 2680 Woodlawn Dr,
Honolulu, HI 96822, USA
\and
LPNHE, Universit\'es Paris VI et Paris VII, 4 place
Jussieu, Tour 33, 75252 Paris Cedex 05, France
\and
School of Physics and Astronomy, 116 Church St. S.E., University of
Minnesota, Minneapolis MN 55455, USA
\and
Department of Physics \& Astronomy, University of British Columbia,
6224 Agricultural Road, Vancouver, BC V6T 1Z1, Canada
}

\date{\today}

\abstract{
We present the first measurement of temperature and polarization
angular power spectra of the diffuse emission of Galactic dust at
353~GHz as seen by Archeops on 20\% of the sky. The temperature
angular power spectrum is compatible with that provided by the
extrapolation to 353~GHz of IRAS and DIRBE maps using \cite{fds} model
number~8. For Galactic latitudes $|b| \geq 5$~deg we report a 4~sigma
detection of large scale ($3\leq \ell \leq 8$)
temperature-polarization cross-correlation $(\ell+1)C_\ell^{TE}/2\pi =
76\pm 21\;\mu\rm{K_{RJ}}^2$ and set upper limits to the $E$ and $B$
modes at $11\;\mu\rm{K_{RJ}}^2$. For Galactic latitudes $|b| \geq
10$~deg, on the same angular scales, we report a 2 sigma detection of
temperature-polarization cross-correlation $(\ell+1)C_\ell^{TE}/2\pi =
24\pm 13\;\mu\rm{K_{RJ}}^2$. These results are then extrapolated to
100~GHz to estimate the contamination in CMB measurements by polarized
diffuse Galactic dust emission. The $TE$ signal is then $1.7\pm0.5$
and $0.5\pm0.3\;\mu\rm{K^2_{CMB}}$ for $|b| \geq 5$ and
10~deg. respectively. The upper limit on $E$ and $B$ becomes
$0.2\;\mu\rm{K^2_{CMB}}\;(2\sigma)$. If polarized dust emission at
higher Galactic latitude cuts is similar to the one we report here,
then dust polarized radiation will be a major foreground for
determining the polarization power spectra of the CMB at high
frequencies above 100~GHz.

\keywords{Cosmic Microwave Background -- Cosmology --
Observations -- Submillimetre -- Polarization -- Dust --
Extinction -- Foreground} }

\authorrunning{N. Ponthieu, J.F. Mac\'{\i}as--P\'erez et al.}
\titlerunning{Dust temperature and polarization power spectra}
\maketitle


\section{Introduction}

The Cosmic Microwave Background (CMB) is now considered as one of the
most sensitive probes to the physics of the early Universe. A great
number of experiments have measured its temperature anisotropy power
spectrum over a wide range of angular scales (for a review, see
\cite{wang}) until WMAP recently gave a cosmic variance limited
estimate up to the rise of the second acoustic peak
(\cite{wmap_cl}).

CMB polarization provides a wealth of complementary information.
First, it allows to break degeneracies that remain in the
determination of cosmological parameters with temperature anisotropy
data alone (see \emph{e.g.}  \cite{zalda_spergel}). Perhaps more
importantly, it sheds light directly on inflation through the $B$ mode
that is generated only by primordial gravity waves produced during
that era (\cite{seljak_97,kamionkowski_97}). Distortions of the $E$
mode by large scales structures weak lensing also induce a $B_{WL}$
signal, distinguishable from the primordial $B$ by its
non-gaussianity. This $B_{WL}$ component in turn provides useful
information about the dark matter distribution ({\it e.g.}
\cite{benabed_01}) and the mass of neutrinos at the precision of
0.04~eV for an experiment e.g. 20 times more sensitive than Planck
(\cite{kaplinghat}).

CMB polarization is however 2 to 5 orders of magnitude smaller than
temperature anisotropies and therefore still remains to be accurately
measured. It is now becoming accessible thanks to improved
instrumental sensitivities. The first detection of the $E$ mode has
been reported by DASI (\cite{dasi_pol}). This result has been
confirmed by the same team (\cite{dasi_04}), CAPMAP
(\cite{barkats}) and CBI (\cite{cbi_04}). WMAP has also provided an
estimate of the $TE$ correlation spectrum
(\cite{wmap_te}) fully compatible with an inflationary scenario.

Instrumental sensitivity is not the only issue in the determination of
the power spectrum of the polarized CMB anisotropies. Other astrophysical components
also contribute to the sky brightness and polarization at the
wavelengths of interest and must be subtracted. These {\sl foregrounds}, 
mainly dominated by the diffuse Galactic emission and
the one of extragalactic point sources, are often not well
constrained or even not experimentally measured in the case of
polarization.

Ground based experiments such as those mentionned above observe small
regions of the sky. They can then choose them where foregrounds are
weak and are less prone to contamination by Galactic emission. It is not
the case for full sky experiments such as WMAP and Planck. At the
Planck--HFI frequencies and for future bolometer experiments, the
dominant component is the radiation from Galactic Interstellar Dust
(ISD). The submillimetre and millimetre (hereafter submm) {\sl
  intensity} of the ISD emission can be inferred from IRAS and COBE--DIRBE
extrapolations ({\sl e.g.}  \cite{fds}, hereafter FDS) and has been measured on large
scales by COBE--FIRAS (\cite{firas1, boulanger, firas2}). However,
little is known about ISD {\sl polarization} emission on scales
larger than 10~arcmin, {\sl i.e.} that are the most relevant for current CMB
studies. Indeed, ground--based observations of submm ISD polarization
are concentrated on high angular resolution (arcminute scale) star
formation regions. Indirect evidence for large scale
polarization comes from the polarization of starlight in extinction
(\cite{serkowski}), but as reviewed by Goodman (1996), these
measurements of background starlight polarization lead to ambiguous
interpretation. In particular, the visible data are biased by low
column density lines of sight and do not fairly sample more heavily
reddened ones. Direct submm measurements are therefore highly
required both for Galactic studies of the large scale coherence of the
magnetic field and in the field of CMB polarization, but are rather
challenging as they require sensitivities comparable to those of CMB
studies.

Recently, \cite{arch_polar} have reported the first measurement of the
submillimetre diffuse polarized emission by interstellar dust in the
vicinity of the Galactic plane using the Archeops experiment. They
show that the Galactic plane is significantly polarized at the 3--5\%
level and that dense clouds can be polarized up to 10\% or more. This
indicates that the dust intrinsic radiation is highly polarized and
that the grain alignment mechanism is very effective.  Considered with
the possible large scale coherence of the polarization orientation, it
shows that the dust polarized emission could be an important
foreground for CMB polarization studies, especially on large angular
scales.

Here, we wish to give a first answer to this question with the
evaluation of the dust polarization power spectra away from the Galactic
plane and in diffuse regions, on angular scales ranging from $\ell =
3$ to $\ell = 70$. Section~\ref{se:instru}
briefly introduces Archeops and its polarization capabilities. Section
\ref{se:processing} presents the processing applied to the data and
Sect.~\ref{se:powerspecestimation} describes the evaluation of the
polarized angular power spectra. Section~\ref{se:results} presents our
main results.  We conclude in Sect.~\ref{se:contamination} with the
extrapolation of our results to lower frequencies where the CMB
dominates, to give an estimate of the dust contamination in the
measurements of the CMB polarization power spectra.

\section{The Archeops instrument \label{se:instru}}

Archeops \footnote{{\tt http://www.archeops.org}.} is a balloon borne
bolometer experiment that aimed at measuring the CMB temperature
anisotropy over large and small angular scales. It provided the first
determination of the $C_\ell$ spectrum from the COBE multipoles
(\cite{smoot}) to the first acoustic peak (\cite{benoit_cl}) from
which it gave a precise determination of the main cosmological
parameters, such as the total density of the Universe and the baryon
fraction (\cite{benoit_params}). Archeops was also designed as a test
bed for Planck--HFI and therefore shared the same technological
design: a Gregorian off--axis telescope with a $1.5\,$m primary
mirror, bolometers operating at common frequencies (143, 217,
353 and 545~GHz) cooled down at $100\,$mK by a $^{3}$He$/^{4}$He dilution
designed to work at zero gravity and similar scanning strategy. A
detailed description of the instrument and its performances can be
found in (\cite{benoit_app}) and (\cite{pipeline_paper}).

At $353$~GHz where dust thermal radiation is dominant, Archeops has 6
bolometers mounted in 3 Ortho Mode Transducer\footnote{Planck-HFI has
since changed for the Polarized Sensitive Bolometer technology to
measure polarization of radiation} (hereafter OMT) pairs that are
sensitive to polarization in order to study the properties of the
polarized dust diffuse radiation. The three OMTs are oriented at
60~degrees from each other to enable the full recovery of the $Q$ and
$U$ Stokes parameters and to minimize the correlations in their
determination. Archeops was launched on February 7th, 2002, from the
CNES/Swedish facility of Esrange, near Kiruna (Sweden). The flight
brought about 12 hours of high quality night data.

\section{Data processing and map making \label{se:processing}}

A detailed description of the data processing and the polarization map
making is given in (\cite{pipeline_paper}) and (\cite{arch_polar}).

The main steps on the Archeops data processing are the followings.
First, the reconstruction of the pointing during flight, with rms error
better than 1~arcmin (\cite{thesealex}), is performed using the data from a bore--sight mounted
optical star sensor aligned to each photometer using Jupiter
observations. Second, corrupted data (including glitches and bursts of
noise) in the Time Ordered Information (TOI), representing less than
1.5\%, are flagged and not considered in the following
processing. Third, low frequency drifts on the data, generally
correlated to house-keeping information (altitude, attitude, cryostat
temperatures, the CMB dipole) are removed using the latter as
templates.  Fourth, high frequency decorrelation is performed in few
chosen frequency intervals of $\sim 1$~Hz width to remove
non-stationary high-frequency noise. Sixth, the corrected timelines
are then deconvolved from the bolometer time constant and the flagged
corrupted data are replaced by a constrained realization of noise.
Finally, low frequency atmospheric residuals and noise are subtracted
using a destriping procedure which preserves the sky signal to better
than 2\% on large angular scales (\cite{thesealex}). To improve the quality of the
removal of atmospheric residuals we have also performed a component
separation analysis in the time ordered data using the SMICA-MCMD
algorithm (\cite{smica}) over all the Archeops channels. From this
analysis we have constructed a template of the atmospheric
contribution to the Archeops data which has been fit and removed
from each of 353 GHz bolometers preserving the dust emission to better
than 5\%.

The six bolometers are cross-calibrated as discussed on Sect.~\ref{se:crosscal}.
The absolute calibration is obtained from a correlation between the
Galactic latitude profiles from FIRAS ``dust spectrum'' maps and those
of Archeops. It has an absolute accuracy of about 12\%. This affects
only the absolute values of $I,\;Q,\;U$ but neither the degree of
polarization nor its orientation. A detailed description of the
calibration is given in \cite{pipeline_paper}.

\begin{figure}[!ht]
\begin{center}
{\includegraphics[clip, angle=0, scale = 0.5]{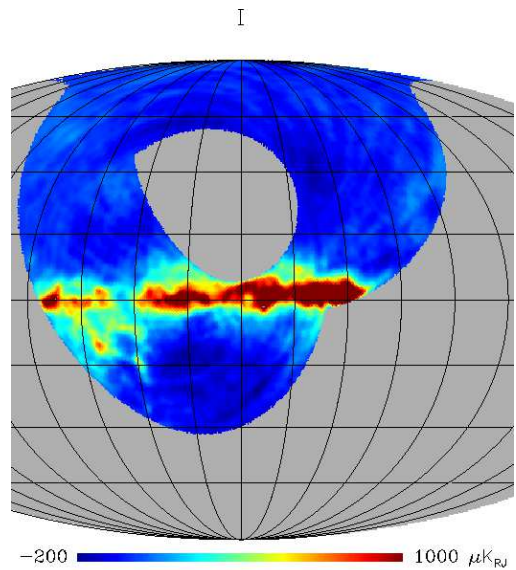}}
\caption{
Total unpolarized intensity $I$ measured by Archeops at 353~GHz. Map
centered on Galactic longitude $l=120$ in Galactic coordinates. The
pixel size is 27~arcmin smoothed by a 2~deg beam FWHM to match the 1.88~deg
pixel size (HEALPix parameter $n_{side} = 32$) used throughout the
analysis. Grid lines are spaced by 20~deg.}
\label{fig:map_i}
\end{center}
\end{figure}

\begin{figure}[!ht]
\begin{center}
{\includegraphics[clip, angle=0, scale = 0.5]{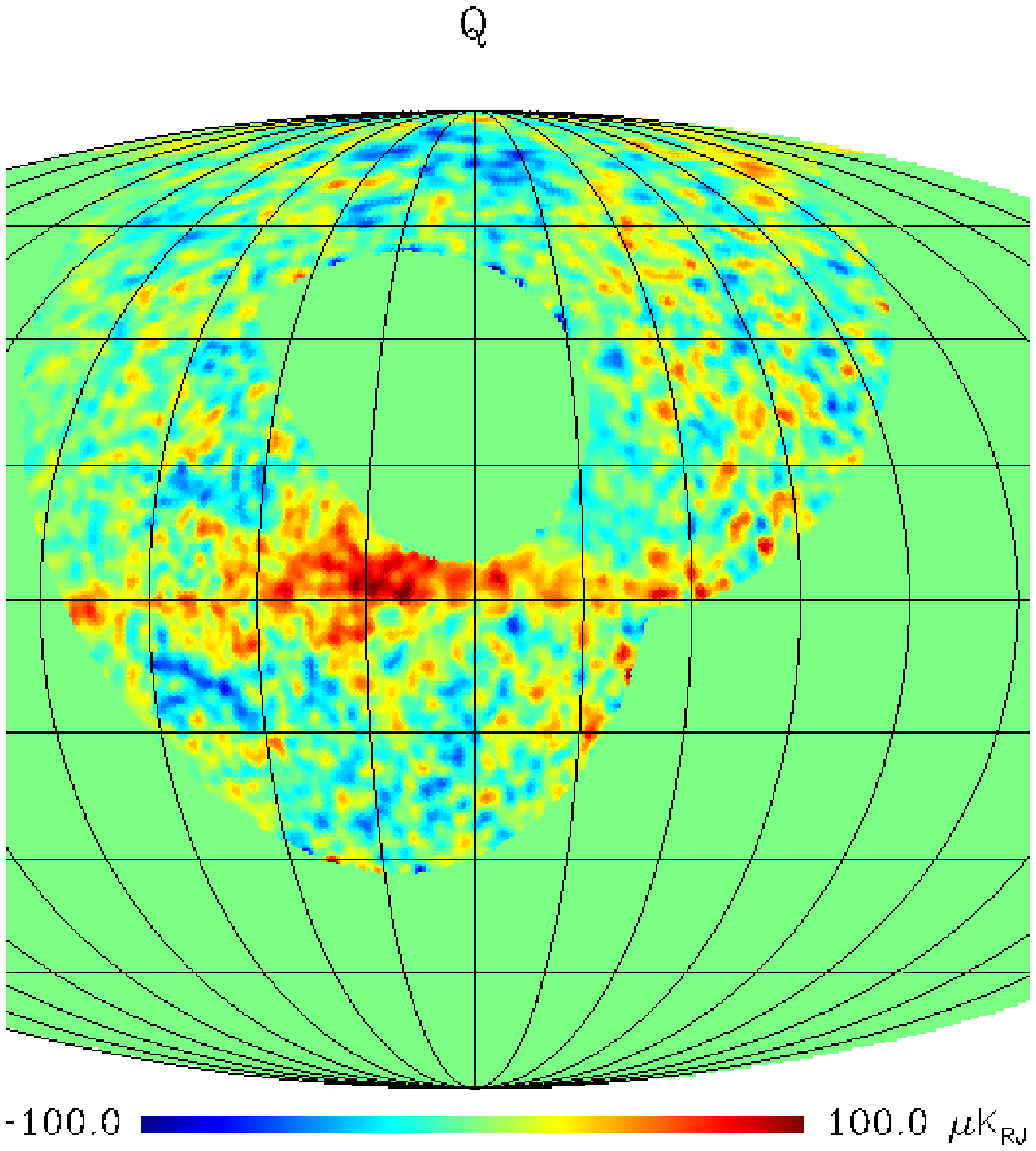}}
\caption{
Stokes parameter $Q$ measured by Archeops at 353~GHz. Map centered on
Galactic longitude $l=120$ in Galactic coordinates. The pixel size is
27~arcmin smoothed by a 2~deg beam FWHM. Grid lines are spaced by 20~deg.}
\label{fig:map_q}
\end{center}
\end{figure}

\begin{figure}[!ht]
\begin{center}
{\includegraphics[clip, angle=0, scale = 0.5]{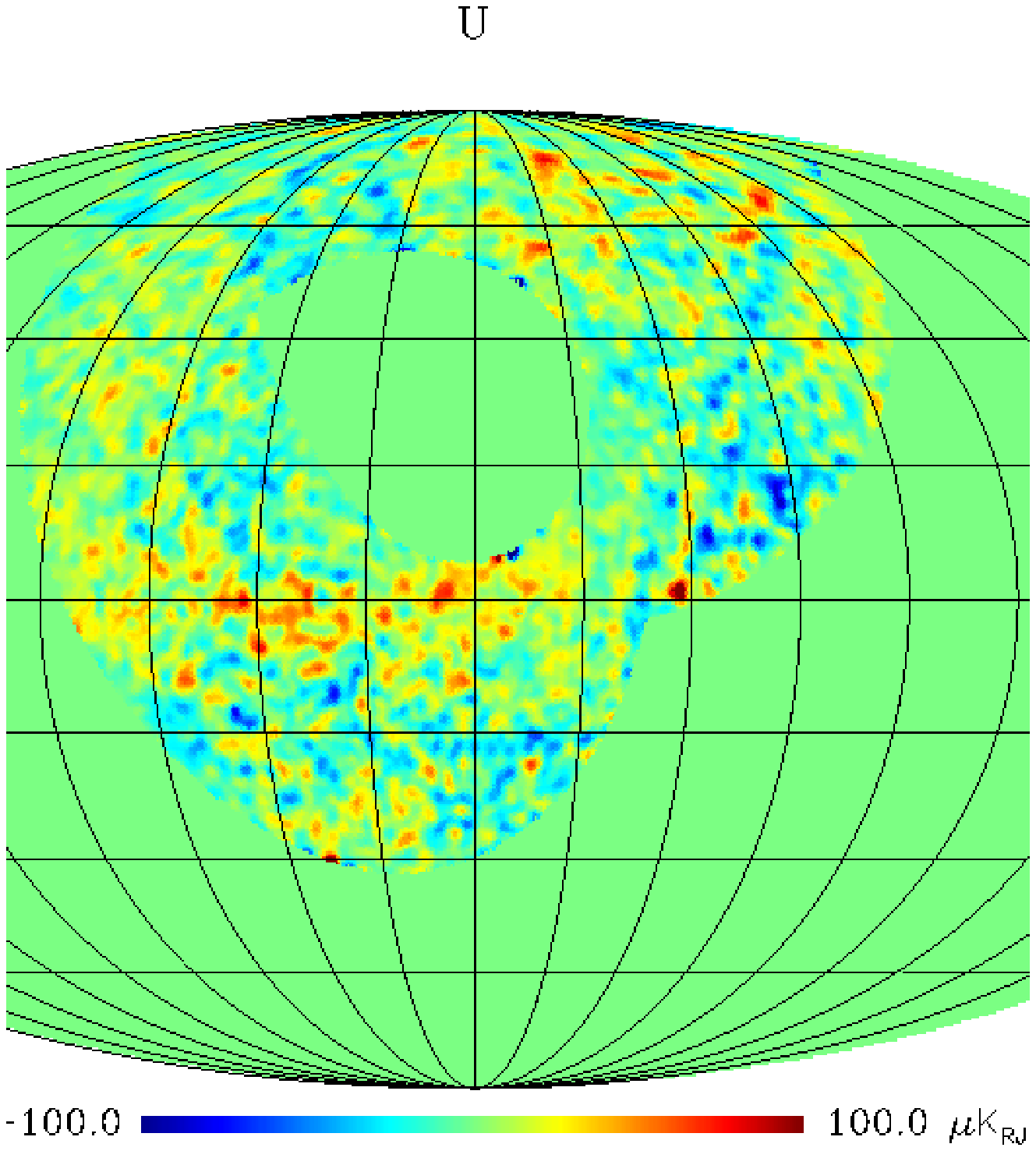}}
\caption{
Stokes parameter $U$ measured by Archeops at 353~GHz. Map centered on
Galactic longitude $l=120$ in Galactic coordinates. The pixel size is
27~arcmin smoothed by a 2~deg beam FHWM. Grid lines are spaced by 20~deg.}
\label{fig:map_u}
\end{center}
\end{figure}

The data processed and calibrated as above have been projected into
polarization maps using the algorithm described in
\cite{ponthieu}. The maps produced are shown on Figs.~\ref{fig:map_i},
\ref{fig:map_q}, \ref{fig:map_u}. A significant polarization level in the Galactic
plane was first reported by Archeops (\cite{arch_polar}) for regions
with longitudes between 100 and 120 degrees and between 180 and 200
degrees. The new data processing allows us to reconstruct polarization
on all the Galactic plane observed by Archeops and to fill the gap
between 120 and 180 degrees.

The noise correlation matrix on those maps have been computed using
pure noise simulations which are described in Sect.~\ref{se:noise}.

\section{Evaluation of the polarization power spectra\label{se:powerspecestimation}}

\subsection{Formalism\label{se:formalism}}

For a direction of observation ${\bf n}$, we define the Stokes
parameters $I$, $Q$ and $U$ in the tangential plane with respect to $({\bf
-e_{\theta}}, {\bf e_{\varphi}})$. The angle of the polarization is
oriented from the North Galactic pole through East to the South
Galactic pole (counterclockwise).

For statistical analysis, the use of $E$ and $B$ (\cite{seljak_97}) is
however more suitable because these quantities are scalar and
independent of the coordinate system.  We here estimate these
quantities and the correlation between $E$ and $T$ using the method
described in \cite{spicepol}. The spin-2 nature of Stokes parameters
leads us to define

\begin{equation}
P \equiv Q + iU.
\end{equation}

When the polarized two point correlation functions are estimated in
real space between two directions $\hat{n}_1$ and $\hat{n}_2$, one has
to rotate $P$ by angles $\alpha_1$ and $\alpha_2$ respectively to
align the axes defining $Q$ for each direction of observation with the
geodesic connecting $\hat{n}_1$ and $\hat{n}_2$ such that

\begin{eqnarray}
\overline{P}(\hat{n}_1) &\equiv& e^{2i\alpha_1}P(\hat{n}_1) \\
\overline{P}(\hat{n}_2) &\equiv& e^{2i\alpha_2}P(\hat{n}_2).
\end{eqnarray}

We then define the correlation functions

\begin{eqnarray}
\xi_{-}(\theta) &\equiv&
\langle \overline{P}(\hat{n}_1)\overline{P}(\hat{n}_2) \rangle \nonumber\\
&=&\sum_{\ell}\frac{2\ell+1}{4\pi}
(C_{\ell}^{EE}-C_{\ell}^{BB})\,d^\ell_{2-2}(\theta)\, , \label{eq:xi_minus}\\
\xi_{+}(\theta) &\equiv&
\langle \overline{P}^*(\hat{n}_1)\overline{P}(\hat{n}_2) \rangle \nonumber\\
&=&\sum_{\ell}\frac{2\ell+1}{4\pi}
(C_{\ell}^{EE}+C_{\ell}^{BB})\,d^\ell_{22}(\theta)\, ,\label{eq:xi_plus}\\
\xi_{\times}(\theta) &\equiv&
\langle T(\hat{n}_1)\overline{P}(\hat{n}_2) \rangle\nonumber\\
&=&\sum_{\ell}\frac{2\ell+1}{4\pi}
C_{\ell}^{TE}\,d^\ell_{20}(\theta)\, ,\label{eq:xi_cross}
\end{eqnarray}

where $\hat{n}_1 \cdot \hat{n}_2 = \cos\theta$ and $d^l_{mm^\prime}$
are the reduced Wigner $D$-matrices. The $C_\ell$s angular power spectra are
then obtained by the following integration:

\begin{eqnarray}
C_\ell^{EE} - C_\ell^B - 2iC_\ell^{EB} & = & 2\pi\int_{-1}^1\hat{\xi}_-
(\theta)d^\ell_{2-2}(\theta)d\cos\theta\, , \label{eq:ctheta2cleb}\\
C_\ell^{EE} + C_\ell^B & = & 2\pi\int_{-1}^1\hat{\xi}_+(\theta)d^\ell_{22}(\theta)d\cos\theta\, ,  \\
C_\ell^{TE} + iC_\ell^{TB} & = & 2\pi\int_{-1}^1\hat{\xi}_
\times(\theta)d^\ell_{20}d\cos\theta\, . \label{eq:ctheta2clte}
\end{eqnarray}

We estimate those quantities with the software SpicePol
(\cite{spicepol}) that uses the HEALPix package (\cite{healpix}) to
compute the pseudo-$C_\ell$s from the raw maps :
$\tilde{C}_\ell^{TE}$, $\tilde{C}_\ell^{EE}$, $\tilde{C}_\ell^{BB}$,
from which we obtain an estimate of the signal plus noise angular
power spectra. The noise contribution is estimated through Monte-Carlo
simulations (Sect.~\ref{se:noise}) and subtracted, producing estimates
of the angular polarized power spectra. Because of approximations used
by SpicePol it can not produce estimates of $\tilde{C}_\ell^{TB}$
and $\tilde{C}_\ell^{EB}$.

The total uncertainty on the determination of the power spectrum
$C_\ell^{TT}$ includes both the variance of the noise and the variance
of the signal itself. In the case of Galactic dust, there is no cosmic
variance {\it stricto sensu}. Indeed, the dust distribution in the sky
and its properties are not a particular realization of a Gaussian
random field like in the case of the CMB anisotropies. Therefore, we
compute the variance of the polarized power spectra considering only
the noise-noise and the signal-noise contributions, and not the
signal-signal covariance (sample variance) as follows

\begin{eqnarray}
\sigma^2(C_\ell^{XX}) = \frac{2}{(2l+1)f_{sky}}&&N_\ell^{XX\,2} \label{eq:sigma_clte1}\\
\sigma^2(C_\ell^{TE}) = \frac{1}{(2l+1)f_{sky}}&&
\left[N_\ell^{TE\,2} + N_\ell^{TT}C_\ell^{EE} \right. \nonumber \\
&& \left. + N_\ell^{EE}C_\ell^{TT} + N_\ell^{TT}N_\ell^{EE}\right] \label{eq:sigma_clte}
\end{eqnarray}

where $X$ stands for $T$, $E$ or $B$, $N$ is the noise and $f_{sky}$
is the fraction of the sky taken for the analysis. The dominant term
in Eq.~(\ref{eq:sigma_clte}) is the product of the covariance of
temperature by the polarization noise $N_\ell^{EE}C_\ell^{TT}$.  These
relations are however only handy approximations to the true
uncertainties and need an empirical adjustment of the $f_{sky}$
parameter that is performed using the simulations described in
Sect.~\ref{se:noise}. These simulations are also used to compute the
noise power spectra.

The temperature power spectrum $C_\ell^{TT}$ and its error bars are
computed using the Xspect method (\cite{xspect}) which uses 15 cross
power spectra from the six detectors and no auto power spectrum in
order to avoid corrections induced by the detector noise.

\subsection{Instrumental noise \label{se:noise}}

The noise power spectrum of the bolometers is estimated from a four
step process already applied in Beno\^\i t~\emph{et~al}~(2003a) in the
context of the $C^{TT}_\ell$ evaluation. First, the Galactic plane region
is masked and interpolated in the time domain with slowly varying
functions. Secondly, these timelines are projected onto maps that are
deprojected to obtain a second timeline with a higher signal to noise
ratio. We then subtract this second timeline from the original one to
obtain a noise dominated timeline and compute its time domain power
spectrum. Simulations including realistic noise and Galactic
contamination show that this process allows the recovery of the true
noise power spectrum at the 5\% level.

From these time domain noise power spectra, we generate noise
timelines for the six polarized bolometers and project them onto maps
in the same way as the real data. The same statistical analysis as the
one applied to the data is performed on 250 of such noise maps
to have a good estimate of the noise angular power spectra. These
power spectra are then subtracted from the Archeops polarized angular
power spectra in order to correct them from the noise bias. Finally,
we compute the uncertainties on the polarization power spectra using
equations (\ref{eq:sigma_clte1}) and (\ref{eq:sigma_clte}).

\subsection{Systematic effects\label{se:syste}}

Three main sources of systematic effects are likely to affect the
evaluation of the polarization correlations: the filtering applied to
the data, the uncertainty on the cross-calibration between the
detectors and the uncertainty on the knowledge of the exact
orientation of the polarizers on the focal plane. We address these
three issues separately.

\subsubsection{Filtering}

A bandpass filtering is applied to the data. The lower frequency bound
is $0.033\,$Hz, which corresponds to the first harmonic $f_{spin}$ of
the rotation of the gondola (2~rpm). Because of the spinning of the
instrument, few physical components at frequencies below $f_{spin}$ in the timeline
are projected on the map. The high pass therefore ensures that
no physically irrelevant and dipole-like components remain in the
timelines. A low pass filtering is then applied to the timelines at
$38\,$Hz to remove high frequency noise. To correct from this effect
on the angular power spectra we have computed an effective filtering
function $F_\ell$ in the multipole space from simulations of
temperature Gaussian fields with a flat power spectrum. The maximum
correction is at low $\ell$ and is less than 2\%.

We have also checked that the filtering did not induce any spurious
polarization such as leaks from total intensity into polarization. For
this we have applied the bandpass filter to simulated timelines,
deprojected from the FDS template at 353GHz, for the six bolometers involved in
the determination of the angular power spectra. Then, we have
reconstructed $I$, $Q$ and $U$ maps for which we have extracted the
temperature and polarization power spectra. No spurious polarization
was produced at the level of 0.1\%.

\subsubsection{Cross calibration\label{se:crosscal}}

Stokes parameters are mainly estimated from the differences of the
outputs of bolometers that measure orthogonal polarization states. An
error in the cross calibration between detectors generates a
systematic leak of total intensity into polarization. The cross
calibration of Archeops channels is described in details in
\cite{arch_polar} and relies on the scaling of Galactic intensity profiles
computed from each bolometer. The cross calibration coefficients are
then determined at the 2\% level. In order to give a conservative
upper limit to the effect of this uncertainty on the angular power
spectra, we forced the cross calibration coefficients to values which
deviate by 2\% and maximized the relative difference between two
orthogonal photometers. The angular power spectra were then estimated
with this new set of coefficients and the error bars were derived from
the standard deviation of 200 simulations, assuming a symmetric
uncertainty. This uncertainty is about 5\% of the statistical error
bars.

\subsubsection{Polarizer relative orientation\label{se:polrelorientation}}

The accurate positioning of the polarizers in the focal plane is made
difficult by the complexity of the instrument. We therefore checked
that they were correctly placed during the pre--flight ground
calibrations and found that they were indeed in their nominal
configuration with a $1\,\sigma$ uncertainty of $3\,$degrees. A
mismatch between the assumed and the real orientations of the
polarizers generates a relative error of a few percents on the
reconstructed Stokes parameters (\cite{kaplan}). In order to estimate
the error induced by the uncertainty of the knowledge of the accurate
positionning of the polarizers, we have performed simulations in which
these angles are forced to random values different from those used for
the $I$, $Q$, $U$ reconstruction. The resulting uncertainty is of the
same order as that of the cross calibration uncertainty, that is to
say about 5\% of the statistical error bars.

\section{Results \label{se:results}}

\begin{figure*}[!ht]
\begin{center}
{\includegraphics[clip, angle=0, scale = 0.4]{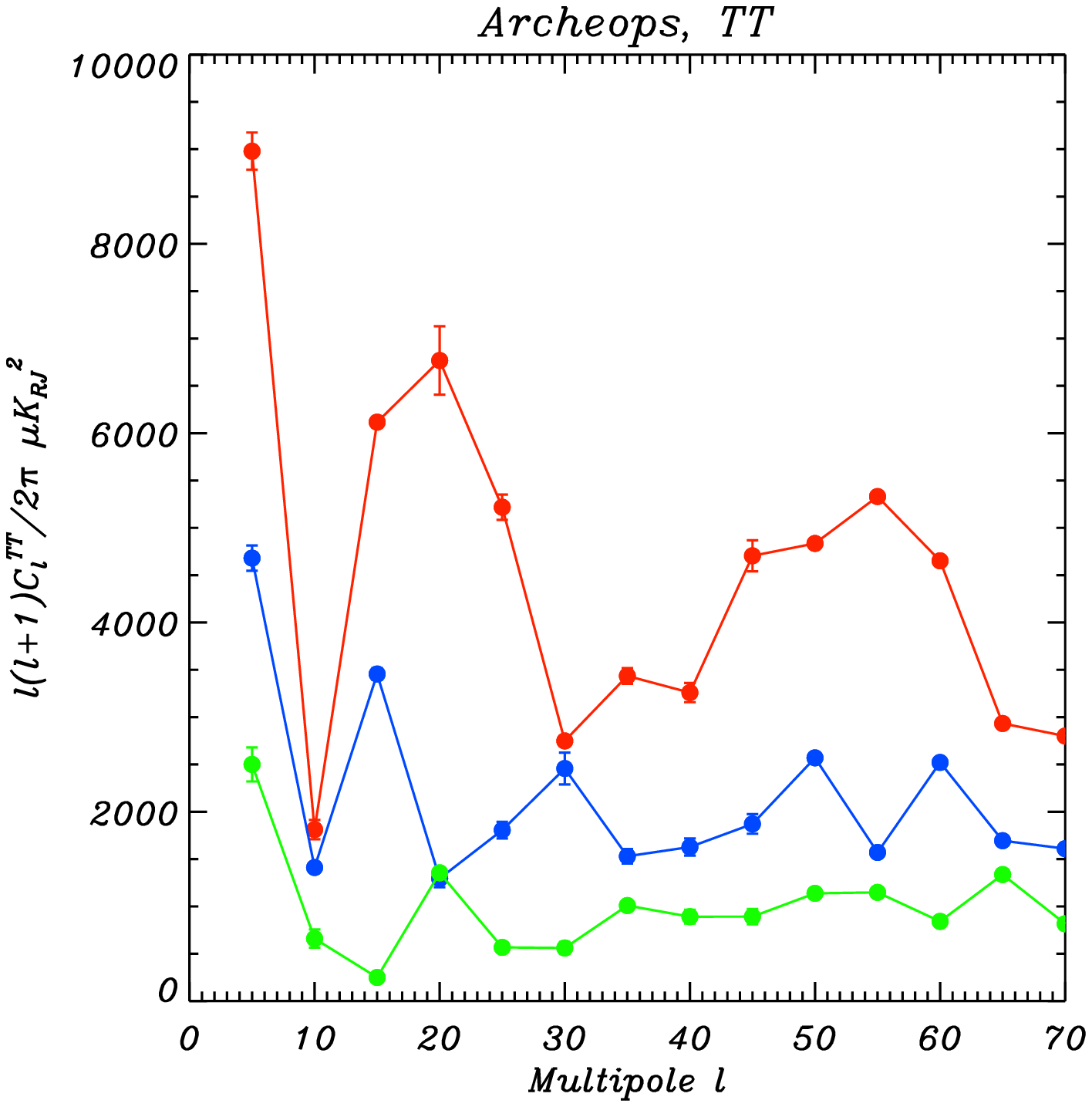}}
{\includegraphics[clip, angle=0, scale = 0.4]{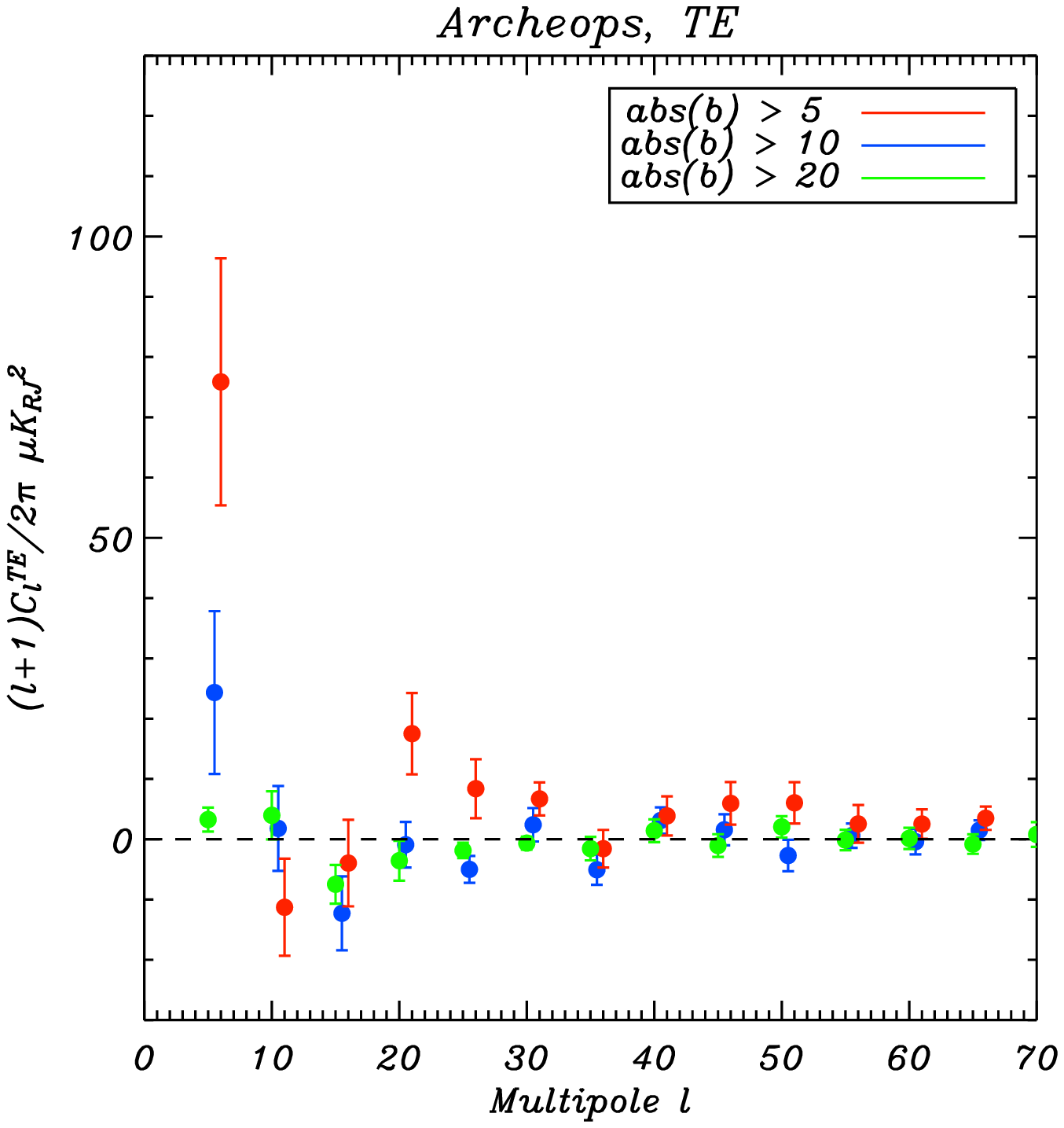}}
{\includegraphics[clip, angle=0, scale = 0.4]{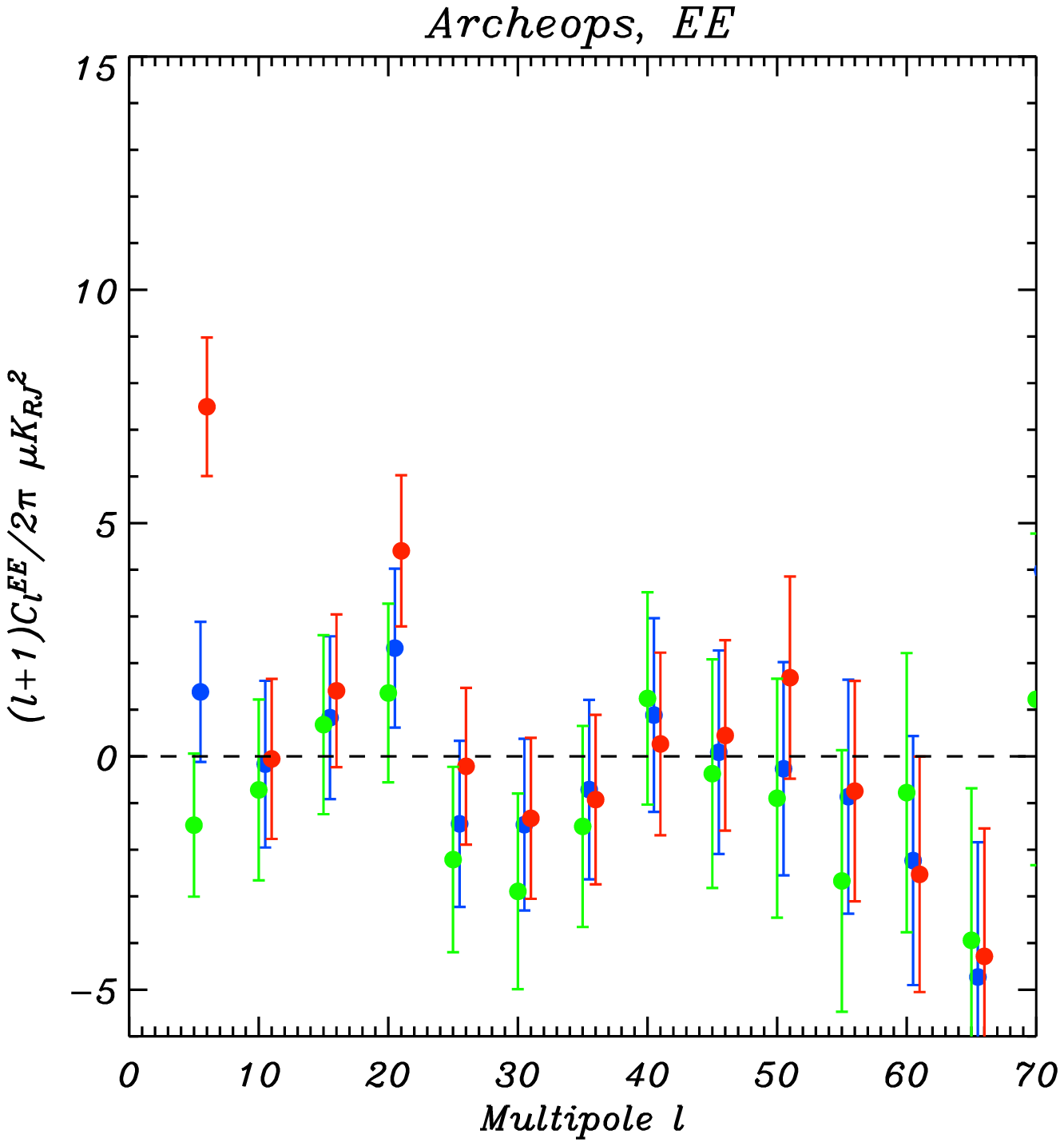}}
{\includegraphics[clip, angle=0, scale = 0.4]{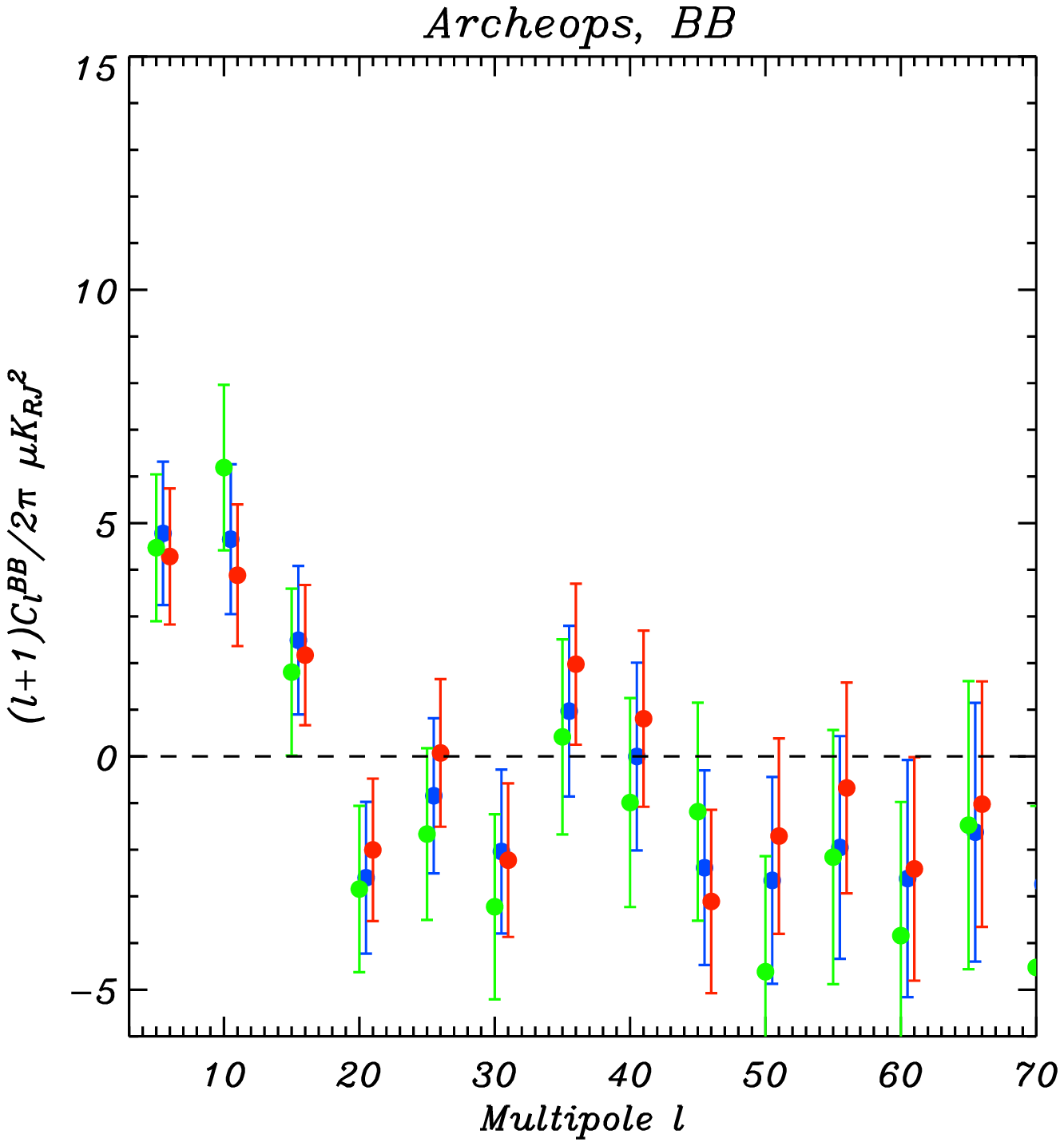}}
\caption{
Clockwise from top left: power spectra $C_\ell^{TT}$, $C_\ell^{TE}$,
$C_\ell^{BB}$ and $C_\ell^{EE}$ computed from the 353~GHz Archeops
data for three different Galactic cuts $|b| \ge 5$, 10, and 20~deg. At
low $\ell$ the power in $C_\ell^{TT}$, $C_\ell^{TE}$, and
$C_\ell^{EE}$ decreases with increasing $|b|$, as would be expected
from a Galactic signal. Since the power in $C_\ell^{BB}$ does not
change with $|b|$ its source is probably not astrophysical. ASCII
files of these data can be obtained at
http://www.archeops.org/info$_{-}$polar.html}
\label{fig:multi_galcuts}
\end{center}
\end{figure*}

We estimate the angular power spectra as a function of cuts of the
data in Galactic latitude in order to remove the effects of regions
with the strongest dust emission along the Galactic plane.
Figure~\ref{fig:multi_galcuts} gives the power spectra for $|b|\geq
5$, $|b| \geq 10$, and $|b|\geq 20$. The spectrum for $C_\ell^{TT}$
oscillates as a function of $\ell$, with the amplitude of oscillation
increasing with decreasing $\ell$. These features are consistent
with a Galactic origin for the signal as discussed in
Sect.~\ref{se:cosecant}.

{\bf The $TE$ Spectrum:} on large angular scales $3\leq\ell\leq8$
there is a $\sim 4\sigma$
detection for $|b|\geq5$ with a magntidue of $(\ell+1)C_\ell^{TE}/2\pi = 76\pm
  21\;\mu\rm{K_{RJ}}^2$ and a $\sim 2\sigma$ detection for
$|b|\geq10$ with a magnitude of $(\ell+1)C_\ell^{TE}/2\pi = 
24\pm 13\;\mu\rm{K_{RJ}}^2$.
For $|b|\geq 5$ there is also a $\sim 2~\sigma$ detection for
$18\leq\ell\leq23$, corresponding to a peak in the temperature
spectrum at the same $\ell$ bin.

{\bf The $EE$ Spectrum:} on large angular scales $3\leq\ell\leq8$
there is a $\sim 5\sigma$ detection for $|b|\geq 5$ with a magnitude
of $(\ell+1)C_\ell^{EE}/2\pi = 7.5\pm 1.5\;\mu\rm{K_{RJ}}^2$ and
a $\sim 2~\sigma$ for
$18\leq\ell\leq23$. Otherwise, the power in the $EE$ spectra is
consistent with zero for all $\ell$'s for $|b|\geq 10$ and for
$|b|\geq 20$ and at $20 \leq \ell \leq 70$ for all latitude cuts.
Similar to the $TE$ spectrum, the power measured for $EE$ on all
scales decreases with increasing latitude, which is consistent with a
Galactic origin for the signal.

{\bf The $BB$ Spectrum:} there are detections of power in the first
few $\ell$ bins and the spectrum is consistent with no power for $\ell
\geq 19$.  The power at the low $\ell$ bins does not depend on
Galactic latitude cut. We argue below that models of polarization from
dust emission predict a decrease of power in both the $EE$ and $BB$
spectra with Galactic latitude cut and therefore the $BB$ signal at
low $\ell$ is probably spurious.  Unlike the CMB for which $E$ and $B$
have different physical sources, dust is expected to produce
comparable amounts of $E$ and $B$. Assuming a similar level of $E$ and
$B$ and interpreting the data in terms of upper limits we provide a
common $2 \sigma$ upper limit to both modes at
$(\ell+1)C_\ell^{EE,BB}/2\pi < 11\;\mu\rm{K_{RJ}}^2 $ at 353~GHz for
$|b|\geq5$.

Since the Archeops data probe the dust diffuse emission and not local
clouds in the Galactic plane it is likely that the orientation and
coherence of the magnetic field is similar for latitudes larger than 5
and larger than 10~deg.  Therefore, the more dust there is along the
line of sight, the more intense the emission should be, and the more
power we expect in the temperature and polarization spectra. According
to these arguments there should be a monotonous decrease of power in
the spectra with increasing latitude cuts, as observed for the $TT$,
$TE$ and $EE$ data.

\subsection{Comparison with models}
In this section we compare our results to models of diffuse Galactic
dust emission. First, we compare the $TT$ power spectrum to those
expected using a Galactic cosecant-law model and using the FDS model.
Second, we obtain an alternative estimate of the TE spectrum by
cross-correlating the FDS template with the Archeops Q and U
maps. Finally, we compare the measured spectra to those calculated on
the basis of a simplified physical model of polarized emission from
dust.

\subsubsection{Pure cosecant-law Galactic emission \label{se:cosecant}}

Due to the disk geometry of the Galaxy, the integrated emission along
a given line of sight increases as the absolute value of the latitude
decreases. This is well approximated by a cosecant law of the
form $I(b) \propto 1/\sin(b)$. When this Galactic contribution
is subtracted, the angular power spectrum of the remaining  
diffuse dust has the form $C_\ell \propto \ell^{-3}$ 
(\cite{gautier,wright}).

\begin{figure*}[!ht]
\begin{center}
{\includegraphics[clip, angle=0, scale = 0.4]{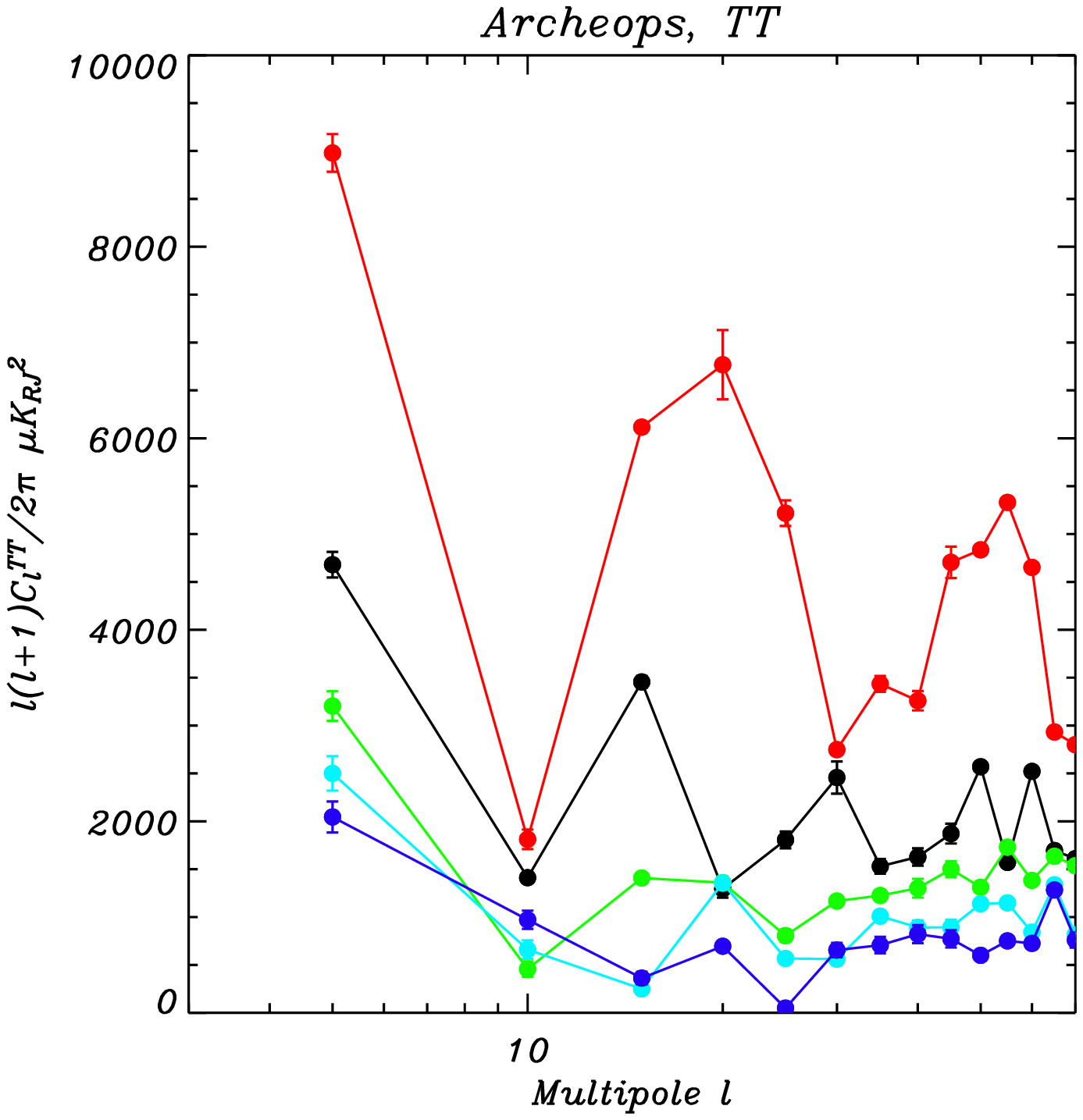}}
{\includegraphics[clip, angle=0, scale = 0.4]{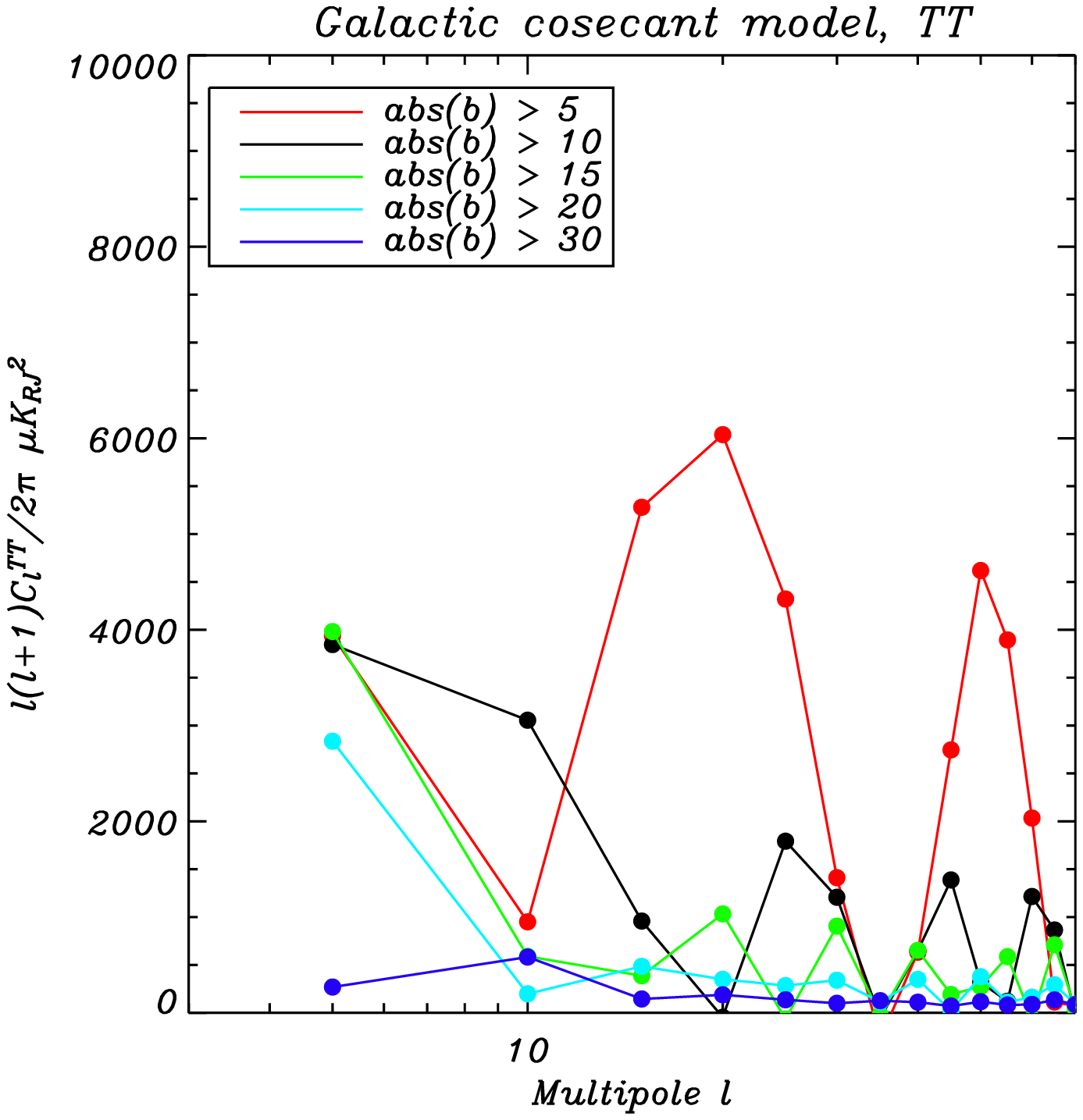}}
{\includegraphics[clip, angle=0, scale = 0.4]{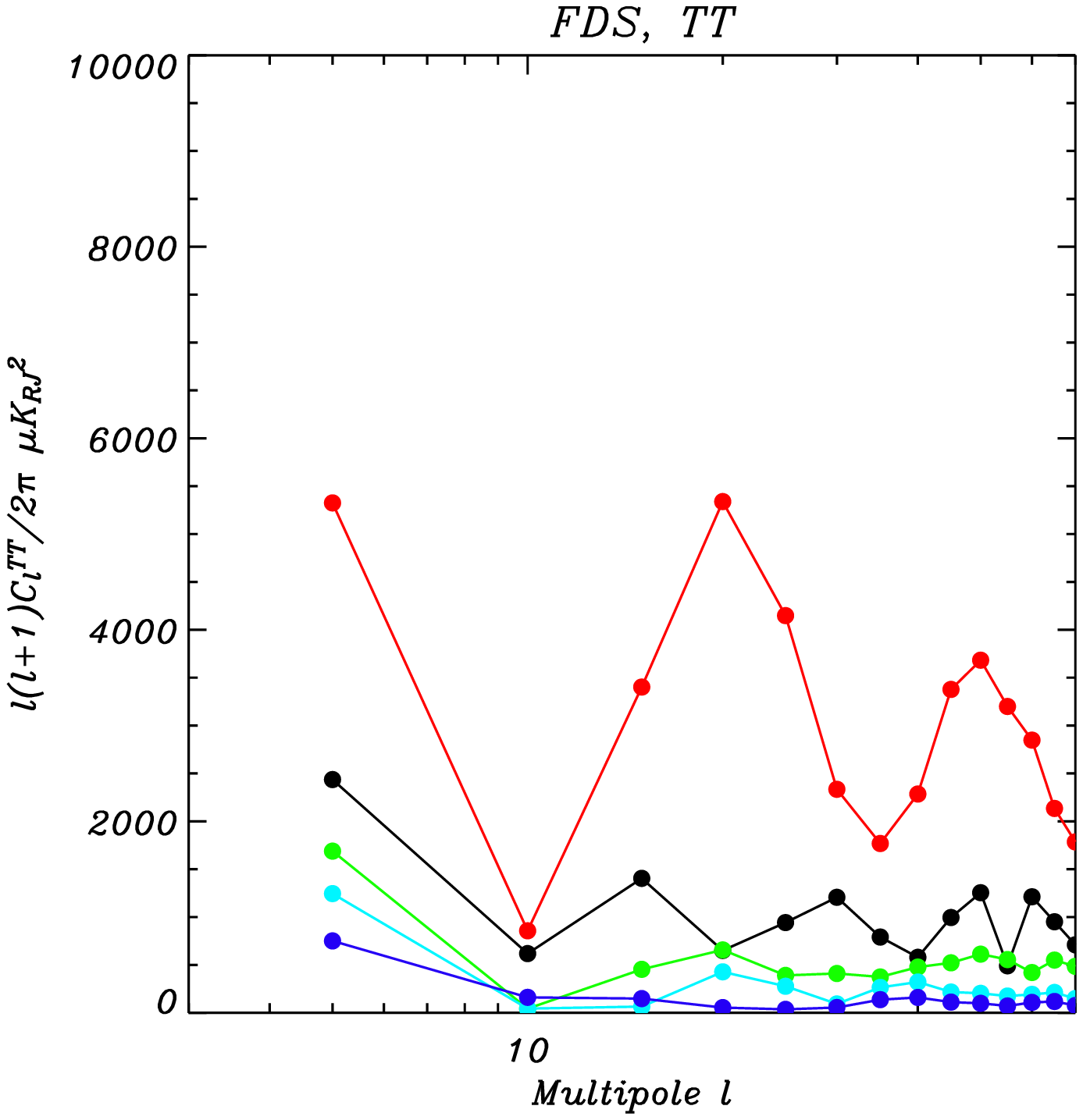}}
\caption{
Temperature angular power spectrum of Archeops data (upper left
diagram), of a pure Galactic cosecant emission law simulation (upper
right diagram) and of a FDS template (bottom diagram) computed using
Xspect for five different Galactic cuts.}
\label{fig:cosecant}
\end{center}
\end{figure*}

We choose to leave the Galactic contribution in our data such that we
can assess its magnitude and its potential contamination of CMB
polarization experiments.  (see Sect.~\ref{se:contamination}). In
order to estimate the influence of a cosecant law component in our
data, we simulate such an emission analytically and with an amplitude
compatible with the FDS template. The results are presented on
Fig.~\ref{fig:cosecant} where we plot the $TT$ angular power spectrum
at different Galactic latitude cuts for the Archeops data, for the
cosecant-law model and for the FDS template. There is a qualitative
agreement between the oscillation pattern of the Archeops data for low
Galactic latitude cuts and the pattern present for the cosecant-law
model and for the FDS template. The agreement suggests that the
cosecant-law emission dominates the observed Galactic dust emission at
large angular scales.

\subsubsection{$FDS$-Archeops correlation \label{se:sfdarch}}

\begin{figure*}[!ht]
\begin{center}
{\includegraphics[clip, angle=0, scale = 0.4]{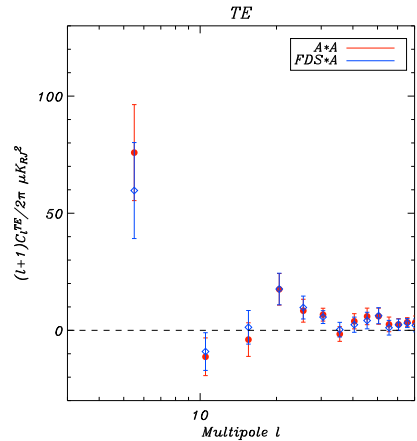}}
\caption{
$TE$ angular power spectra for $|b|\geq5$ either with Archeops data alone 
(red, circles) or with Archeops $Q$ and $U$ data and $I$
from the FDS template (blue, diamonds) on the Archeops region 
of observation extrapolated to 353~GHz (\cite{fds}).}
\label{fig:arch_sfd_te}
\end{center}
\end{figure*}

It is interesting to correlate the Archeops $Q$ and $U$ data with
available dust intensity maps. We quantify the correlation by
calculating the cross-spectrum between the two data sets. This idea
has already been succesfully applied to temperature anisotropy data
sets (\cite{abroe,tristram}.  For dust intensity we use the FDS
template, based on IRAS $100~\mu$m maps and extrapolated to 353~GHz
(``model number 8'') (\cite{fds}). We found that this model fits the
Archeops temperature data at 353~GHz with fractional deviations of
less than 20\%.  Figure~\ref{fig:arch_sfd_te} shows that the
``FDS$\times$Archeops'' cross-spectrum (in blue) and the
``Archeops$\times$Archeops'' $TE$ power spectrum (in red) are
consistent within the error bars. This result provides additional
confidence that the detection of power in the Archeops$\times$Archeops
$TE$ spectrum at low $\ell$ is due to polarized dust and is not
spurious.  We note that the cross-spectrum with the FDS map avoids
correlated noise that is inherent in the Archeops$\times$Archeops
spectrum and thus the agreement between the two results suggests that
correlated noise did not induce spurious results.

\subsubsection{Comparison to a simple physical model of Galactic dust 
polarization\label{se:bspiral}}

In a given direction of observation, the measured Stokes parameters
are the result of the integral along the line of sight of the local
Stokes parameters. These, in turn, depend on the local alignment of
dust grains with the magnetic field and on the intrinsic degree of
polarization. Precise modelling of the diffuse emission due to dust
and its polarized angular power spectra is a complex problem. Previous
work has shown that on large scales, the alignement of dust grains was
compatible with a Galactic magnetic field aligned with the spiral arms
(\cite{fosalba,arch_polar} and references therein). However, such
work relate to data at low Galactic latitudes $|b| \leq 10$~deg, and the
extrapolation at higher latitudes yet remains uncertain due to lack of
data. We here use a toy model of extrapolation at high latitudes of
these polarization properties that gives qualitative agreement with
the Archeops data.

\begin{figure*}[!ht]
\begin{center}
{\includegraphics[clip, angle=0, scale =0.4]{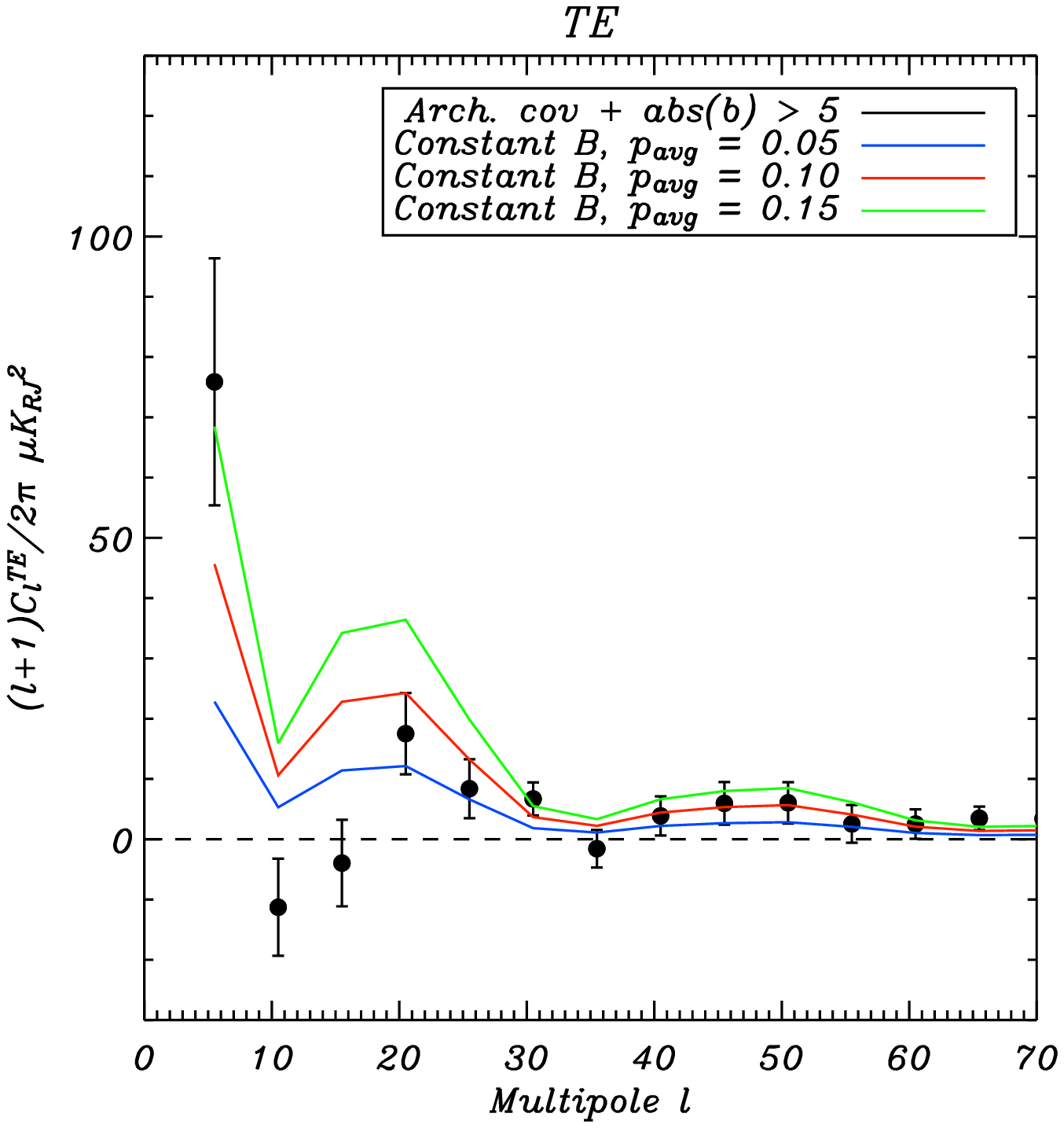}}
{\includegraphics[clip, angle=0, scale =0.4]{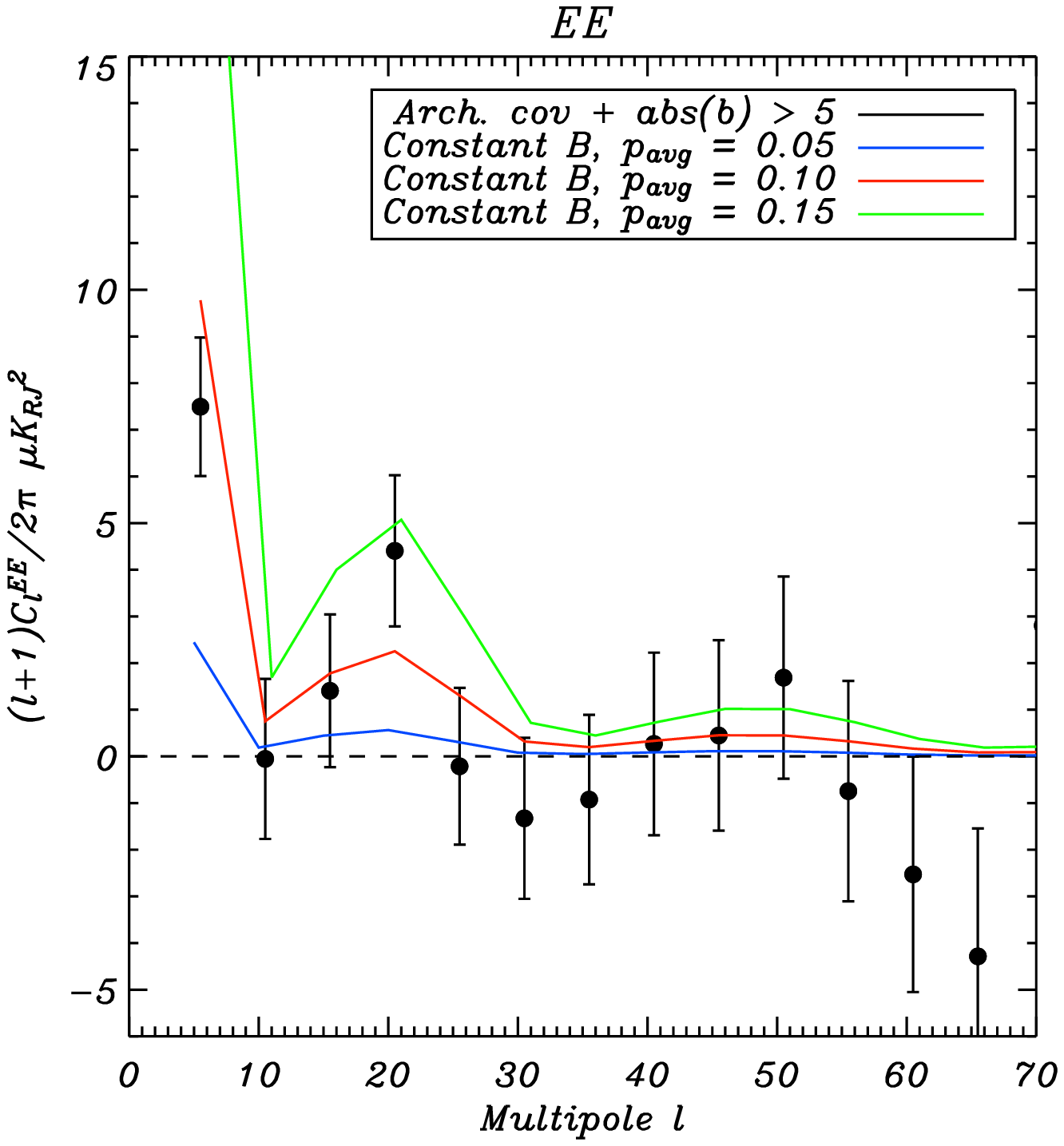}}
{\includegraphics[clip, angle=0, scale =0.4]{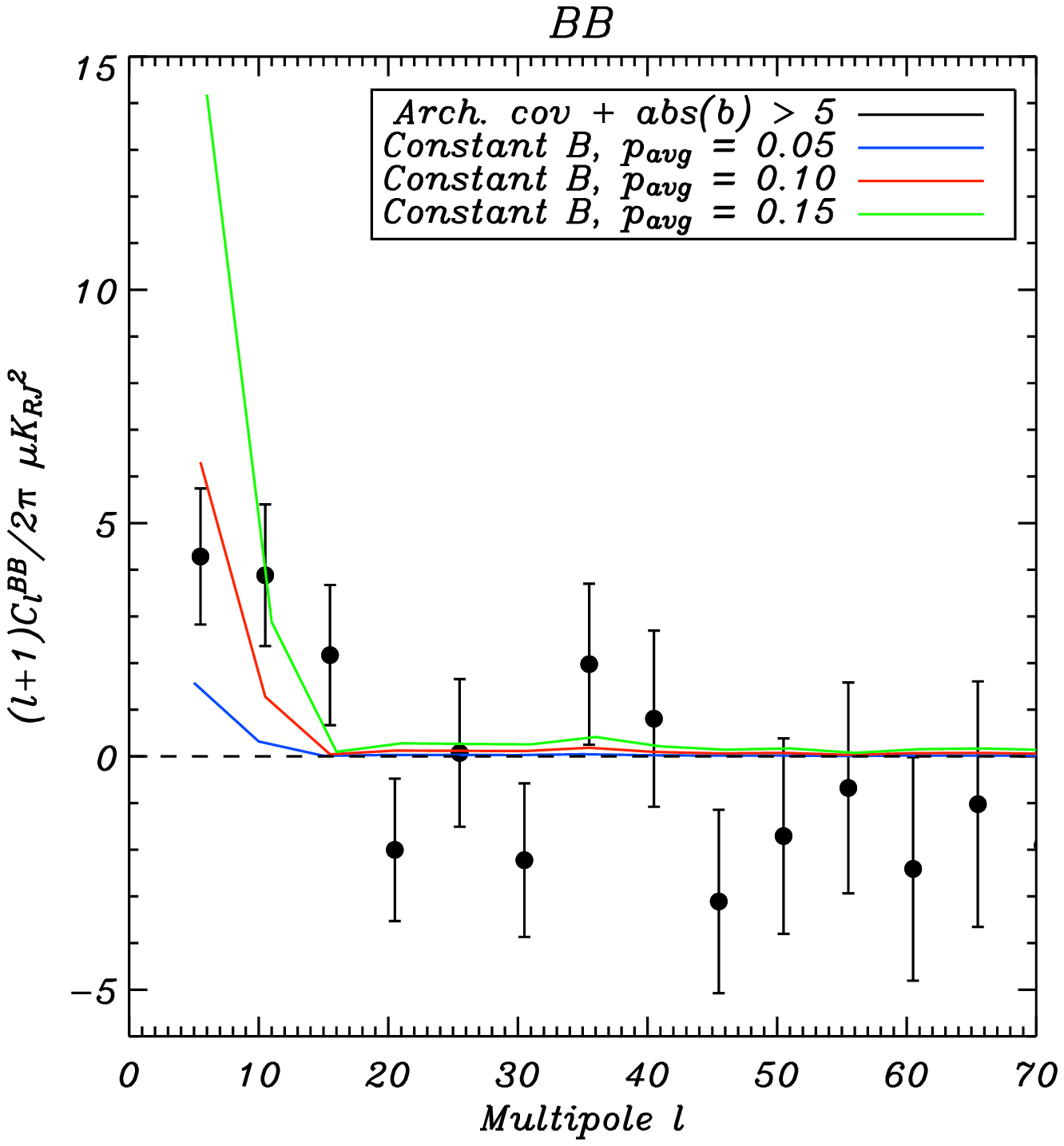}}
\caption{
Comparison between Archeops polarized angular power spectra and those
of a simple model of the Galactic magnetic field for all latitudes
$|b| \geq 5$. The large scale magnetic field is assumed to be
constant and oriented along the local spiral arm. Dust is assumed to
be aligned orthogonaly to it. The effective degree of alignment of the
grains together with the dust intrinsic polarized emissivitiy is
assumed to lead to 5~(blue), 10~(red) and 15\%~(green) effective
degree of polarization.}
\label{fig:compa_archeops_bspiral}
\end{center}
\end{figure*}

Since we are here looking at latitudes away from the Galactic plane,
most of the dust that is being probed is in our vicinity. Typically,
if we take the Galactic disk to be 200~pc thick, a line of sight at $b
= 5$~deg exits the plane at $\sim 1.1$~kpc from the observer. This is
small compared to the size of the spiral arm in which we are located,
and we therefore assume that the large scale magnetic field component
aligned with the local spiral arm is along a constant direction on
this portion of space. A realistic model would of course consider
turbulence, but we are only interested here in a first order
approximation. Based on \cite{arch_polar} who showed that dust diffuse
emission in the vicinity of the Galactic plane was polarized at the
5\% level, and that some dense clouds were polarized at more than
10\%, we present in Fig.~\ref{fig:compa_archeops_bspiral} results with
three assumptions for the level of polarization of Galactic dust:
$p(x) = 5, 10,$ and 15\%. We have only considered latitude cuts of $|b| \geq
5$~deg. Given the simplicity of the model, the relative agreement
between the data and the shape and amplitude of the models is
encouraging.

\section{Dust contamination in CMB polarization estimates\label{se:contamination}}

Our results can be used to estimate the contamination of dust to CMB
polarized angular power spectra. For this we need to extrapolate our
measurements at 353~GHz to lower frequencies where the CMB is usually
measured since it is where it is more intense.

Note that this extrapolation concerns large dust grains only. It does
not take into account the so-called ``anomalous emission'', since (1)
the latter does not contribute significantly at 353~GHz and (2) it is likely
not to be polarized (e.g.~\cite{prunet_laz}). Dust thermal radiation
is the dominant foreground at frequencies above $\simeq 90$~GHz. At
high frequencies it has long been described by a single grey body at a
temperature of 17.5~K with a $\nu^2$ emissivity (\cite{boulanger}). At
low frequencies ($\sim 300$ to $\sim 100$~GHz), the spectrum is not
well constrained due to the lack of accurate measurements. We need to
use the FIRAS and WMAP data jointly. The FIRAS data are used to
compute the intensity value at 353~GHz and the WMAP data give the
intensity value at 94~GHz. This then allows to derive the dust
radiation spectral index.

To determine these dust intensities we use the method described in
Lagache~(2003) and derive the spectrum of the dust whose emission is
correlated with the HI gas, using FIRAS and WMAP data in the Archeops
region. We obtain a ratio $I_\nu(353)/I_\nu(94) = 134$ corresponding
to a spectral index of 1.7 that we use in the following. Note that
this value is very stable when different (large) regions of the sky are
averaged. This is also the same value as the one obtained by
Finkbeiner~\emph{et~al}~(1999) in this frequency range (see discussion
in \cite{fink_04}). Using this spectral index of 1.7, we find for the
largest angular scales ($3\leq\ell\leq8$), $(\ell+1)C_\ell^{TE}/2\pi =
1.7 \pm 0.5$ and $0.5 \pm 0.3\;\mu{\rm K_{CMB}^2}$ for $|b| \geq 5$
and 10~deg. respectively at 100~GHz.

These results suggest that dust may be a very significant foreground
for measurements of the CMB polarization angular power spectra,
particularly if they include Galactic latitudes below $|b| <
10$~degrees.  However, these measurements must be interpreted with
care when considering the entire sky.  Our data includes only $20\%$
of the sky and generalization to the entire sky is questionable
because of the complexity of the statistical properties of dust.

\begin{figure}[!ht]
\begin{center}
{\includegraphics[clip, angle=0, scale =0.4]{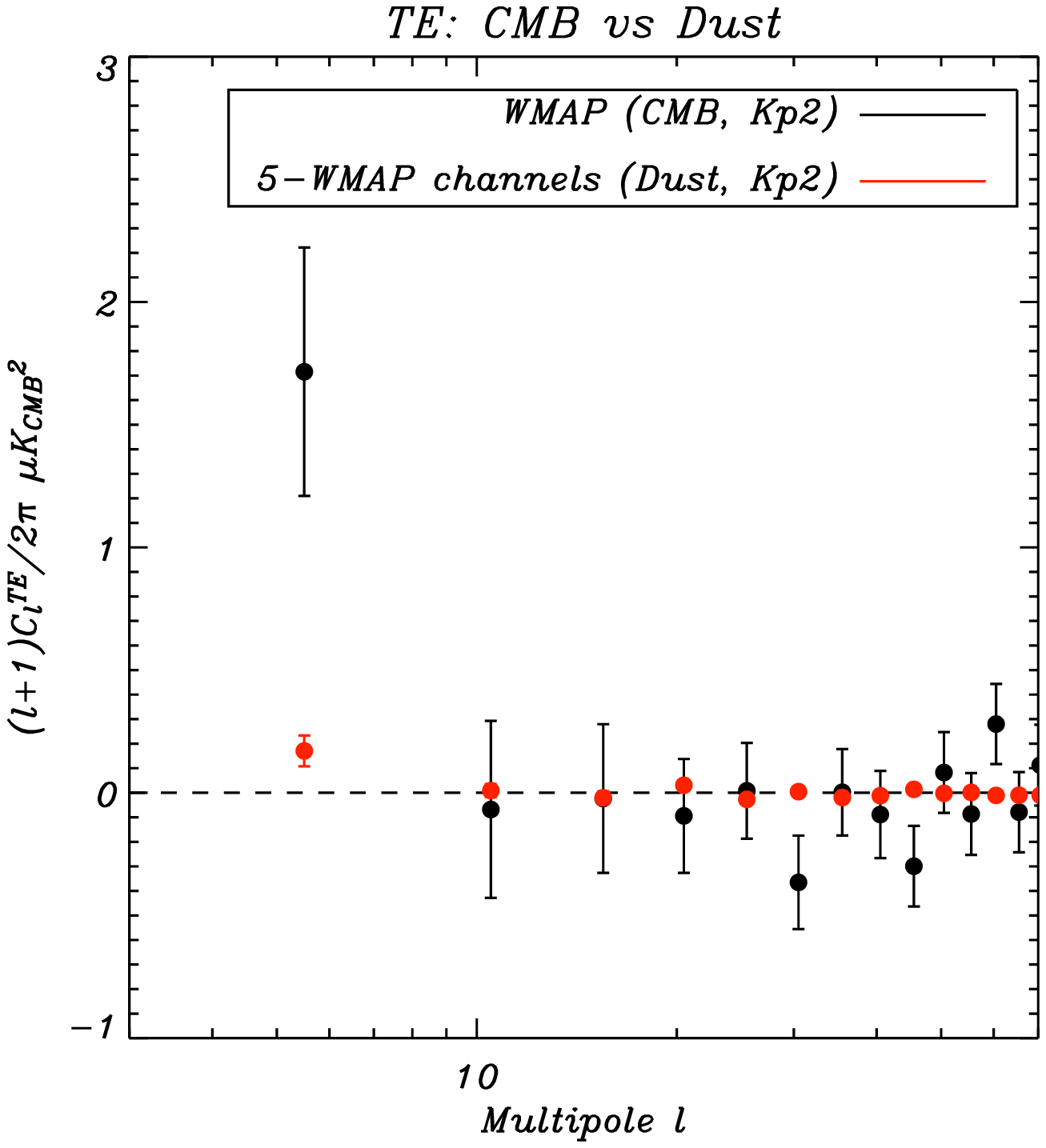}}
\caption{
Dust $TE$ angular power spectrum in CMB temperature units as measured
by Archeops at 353~GHz on the intersection of WMAP's Kp2 mask and the
Archeops sky coverage, extrapolated down to the 5 WMAP frequency
bands 23, 33, 41, 61 and 94~GHz (red) with a constant spectral index
of 1.7 and averaged with an equal weight for each frequency (see
Sect.~\ref{se:contamination} for details). WMAP's
data (first year results) have been rebinned to match
Archeops binning.}
\label{fig:cl_te_100G_guilaine}
\end{center}
\end{figure}

It is interesting to consider the relation between our results and
those of the WMAP team. By combining data from {\it five} frequency
bands at {\it lower frequencies} (23, 33, 41, 61 and 94~GHz) and for
{\it the entire sky} the WMAP team has computed a value for the $TE$
spectrum at $3\leq\ell\leq8$ of
$1.72\pm0.50\;\mu\rm{K^2_{CMB}}$. Accurate extrapolation of the
Archeops results into the WMAP data is complicated and is beyond the
scope of this paper. (For example, it would require a detailed
knowledge of the component separation methods, the relative weighting
of the frequency bands and the shape of the beam). However, it is easy
to {\it illustrate} that combining data at lower frequencies reduces
the effects of dust substantially.  We extrapolate our results to the
WMAP frequency bands with a constant emissivity spectral index of 1.7
in antenna temperature.  We use an equal weight per frequency and use
only the area of the sky that overlaps the Archeops data and WMAP's
Kp2 mask. The result of this extrapolation is presented on
Fig.~\ref{fig:cl_te_100G_guilaine}. On scales of $3\leq\ell\leq8$ we
find $(\ell+1)C_\ell^{TE}/2\pi = 0.17\pm0.06\;\mu\rm{K^2_{CMB}}$,
which is about a factor of 10 smaller than both our estimate at
100~GHz, and WMAP's reported result.

When extrapolated to 100~GHz the upper limits that we found on the $E$
and $B$ modes for $|b|\geq 5$ become $0.2\;\mu\rm{K^2_{CMB}}$.  We
note that the CMB $B$ mode is expected to be at most $\sim
10^{-3}\;\mu\rm{K^2_{CMB}}$ at $\ell \simeq 5$, if the reionization
optical depth is as high as $\tau = 0.17$ (\cite{wmap_params}) and if
the tensor to scalar ratio $T/S$ is the highest compatible with
current CMB temperature anisotropy measurements. If the actual level
of polarized dust over most of the sky is close to the upper limit we
found in this work then a subtraction of a large foreground signal
will be necessary even at 100~GHz in order to detect the primordial
gravitational waves on large angular scales.

\section{Conclusions}

In this paper we have presented the first measurements of the angular
power spectra of the Galactic dust polarized diffuse emission on
approximately 20\% of the sky at 353~GHz by Archeops. On angular
scales $3\leq\ell\leq8$, we obtain a $4~\sigma$ detection of
$(\ell+1)C_\ell^{TE}(dust)/2\pi = 76\pm21\;\mu\rm{K_{RJ}}^2$ for
$|b|\geq5$~deg. On the same angular scales and for $|b|\geq10$, we
have a $2~\sigma$ detection $(\ell+1)C_\ell^{TE}(dust)/2\pi = 24\pm
13\;\mu\rm{K_{RJ}}^2$. This decrease in power is expected from the
cosecant behaviour of the large scale component of the total
intensity, which is shown to dominate the total intensity angular
power spectrum.

On the same sky coverage and for all angular scales $3\leq\ell\leq70$,
we set upper limits to the $E$ and $B$ polarization at
$(\ell+1)C_\ell^{EE,BB}(dust)/2\pi \leq 11\;\mu\rm{K_{RJ}}^2$.
These results have been confirmed by using a template of the Galactic
dust intensity from Finkbeiner~\emph{et~al}~(1999) in place of the
Archeops total intensity map to compute the $TE$ angular power
spectrum. This spectrum agrees with the one derived from Archeops data
alone on all angular scales within $1~\sigma$.

Furthermore, the high degree of polarization seen in the Galactic
plane by Beno\^\i t~\emph{et~al}~(2004a) together with a simple model
of the Galactic magnetic field and the alignment of dust grains, leads
to estimates that are compatible with the data.

To estimate the contribution of Galactic dust to the measurement of
polarized CMB anisotropies, we have extrapolated our results to the
reference frequency 100~GHz, using a spectral index of 1.7, derived
from FIRAS and WMAP data on the Archeops sky coverage. The $TE$ mode
becomes $(\ell+1)C_\ell^{TE}(dust)/2\pi = 1.7\pm0.5$ and
$0.5\pm0.3\;\mu\rm{K^2_{CMB}}$ on $3\leq \ell \leq 8$ for $|b| \geq 5$
and 10~deg respectively. The upper limit on the $E$ and $B$ modes
becomes $0.2\;\mu\rm{K^2_{CMB}}$. These values show that even at
100~GHz where dust radiation is expected to be lower than the CMB, its
polarization may be very significant compared to the CMB and should
be subtracted with care from the observations.

The effects of the polarization of dust are less severe at 
low frequencies. When extrapolating our measurements at 353~GHz
to 20~GHz with the constant spectral index 1.7 and weighting 
the 5 WMAP frequencies equally we find a level of 
$(\ell+1)C_\ell^{TE}(dust)/2\pi =0.17\pm0.06\;\mu\rm{K^2_{CMB}}$
for $3\leq \ell \leq 8$ on the intersection between WMAP's Kp2 mask
and Archeops sky coverage. This is about a factor of 10 smaller
then the WMAP team results.

The high level of polarization measured at 353~GHz by Archeops and
anticipated at the reference frequency of 100~GHz, together with the
uncertainties on dust spectral index and the extrapolation of its
statistical properties to the whole sky call for further precise
studies in order to subtract it precisely from CMB data. This is even
more critical for the detection of the imprint of the primordial
gravitational waves on the CMB $B$ mode, that is expected to be much
smaller than the present upper limit on dust $B$ polarization.

\begin{acknowledgements}
We would like to pay tribute to the memory of Pierre Faucon who led
the CNES team on this successful flight. The HEALPix package was used
throughout the data analysis~\cite{healpix}. We also thank
A.~Starobinsky for fruitful discussions.
\end{acknowledgements}


\end{document}